\documentstyle[12pt]{article}
\begin{document}
\begin{flushright}
Preprint IHEP 93-58 \\
Received 22 April, 1993
\end{flushright}
\begin{center}
Production of four heavy quarks and \\
$B_c$-mesons at the $Z^0$-boson pole\\
\vspace*{1cm}
V.V.~Kiselev, A.K.~Likhoded, M.V.~Shevlyagin\\
\vspace*{4mm}
Institute for High Energy Physics,\\
Protvino, 142284, Russia.\\
E-mail address: Likhoded@mx.ihep.su
\end{center}

\begin{abstract}
The cross-section for the production of $b{\bar b} c{\bar c}$ quarks in
$e^+e^-$ annihilation, that proves to be at a level of
$\sigma (e^+e^- \rightarrow b{\bar b} c{\bar c})/
 \sigma (e^+e^- \rightarrow b{\bar b}) \sim 10^{-2}$ for ${\sqrt s}=M_Z$
is calculated within the frames of the QCD perturbation theory. The cross
sections for the associated production of $1S$- and $2S$-wave states of
$B_c$-meson in the reaction  $e^+e^- \rightarrow B_c b{\bar c}$ were
calculated in the nonrelativistic model of a heavy quarkonium.  The
fragmentation function of $b\to B_c^{(*)}$ is analysed in the scaling limit.
The number of $\Lambda_{bc}$-hyperons to be expected at LEP is estimated
on the basis of the assumption on quark-hadron duality.
\end{abstract}
\newpage

\section*{Introduction.}
The study of heavy quark production processes is of great interest
because this can give  an additional check
of the applicability of the QCD theory to the
description of both free and bound states. The properties of 2- and
3-jet events have been studied   fairly thoroughly at the PETRA, KEK
and LEP colliders. In the light of a high luminosity expected at
the LEP collider, $\sim 10^7 Z^0 $, the study of other rarer processes
is also interesting. Among such processes is the production of four
quarks,  $b{\bar b}c{\bar c}$, in  $e^+e^-$annihilation. We are treating
this process, on the one hand, as verification of the QCD theory  in
higher orders of perturbation theory, $ O({\alpha}^2_s )$,
and, on the other hand,  as a
further step in our understanding of, for example, bound states
$b{\bar c}$ (${\bar b}c)$ not yet discovered experimentally.
The bound states of $b$- and $\bar c$-quarks -- $B_c$-mesons
occupy in the mass spectrum an intermediate state between the families of
$J/\psi$- and  $\Upsilon$-mesons. In this respect the predictions
of the nonrelativistic quark model describing the $J/\psi$- and
$\Upsilon$-families very well can also be applicable for $B_c$-mesons.
Therefore the discovery and study of $B_c$-mesons will become a good
test of QCD and QCD-inspired nonrelativistic potential model.
The processes of $B_c$-meson production in $e^+e^-$-, $p{\bar p}$-
and also neutrino-hadron collisions have been  studied in papers \cite{1}
- \cite{5}.
As to the $B_c$-meson spectroscopy, it was investigated in \cite{6}.
Various aspects
related to  $B_c$-meson decays were considered in \cite{7} - \cite{9}.

Using the exact formulas  of QCD perturbation theory
we calculate the cross-section
$\sigma (¥^+e^- \rightarrow b{\bar b} c{\bar c})$
and also the cross section for $B_c$-meson production in
the process $e^+e^- \rightarrow B_c{\bar b}c$ in nonrelativistic
approximation for $B_c$. We treat these two processes in one work because they
are closely related to each other. The  very convenient method used
to calculate the helicity amplitudes for the process
$¥^+e^- \rightarrow b{\bar b} c{\bar c}$ can also be applied actually
in the same form for calculating the  process
$e^+e^- \rightarrow B_c{\bar b}c$. If one considers the approximate
quark-hadron duality to be valid then the cross section for the
process $¥^+e^- \rightarrow b{\bar b} c{\bar c}$ can be related in the
region of small invariant masses of a $(b{\bar c})$-pair with that
for the production of bound states of $B_c$-mesons in the
process $e^+e^- \rightarrow B_c{\bar b}c$. Using the same
assumption on quark-hadron duality one can also make certain conclusions
on the specific value of the cross-section for the production of
$\Lambda_{bc}$-hyperons -- a bound system of two heavy quarks and one light
valent quark of different flavours.

The paper is arranged as follows. The method of calculating the cross
section for the process $e^+e^- \rightarrow b{\bar b} c{\bar c}$ is
presented in Section 1. Section 2 analyzes the most interesting distributions
for this process. The cross-section for the $B_c$-mesons production
in the process $e^+e^- \rightarrow B_c{\bar b}c$ is discussed in Section 3.
Section 4 presents a brief review of the present experimental situation
regarding the possibility of search for bound states with heavy quarks of
different flavours.
\section{Calculational technique}

At the tree level, 8 Feynmann diagrams, 4 with photon  and 4 with
$Z^0$-boson exchange in the $s$-channel (see Fig.1)
give contribution into the process  of  the
production of two pairs of heavy quarks, $b{\bar b}$ and $c{\bar c}$:
\begin{center}
\begin{equation}
e^+(q_1)+e^-(q_2)\rightarrow b(p_1)+{\bar b}(p_2)+
c(p_3)+{\bar c}(p_4)
\end{equation}
\end{center}

The traditional techniques of calculating these diagrams by the
analytical quadrature  of the amplitude result in too cumbersome
expressions. But still, there are some papers devoted to calculations
of the cross-sections with four-quark final states. For example,
the authors of \cite{10} have studied within the frames of QCD perturbation
theory the cross-section for  the process
$gg \rightarrow q{\bar q} q'{\bar q}'$
for the case of massless quarks and those of \cite{11} have studied it
also for massive quarks.  In \cite{12} the cross-section of the production of
four
quarks in $e^+e^-$-annihilation,  $e^+e^- \rightarrow q{\bar q} q'{\bar q}'$,
for the case of massless quarks
has been calculated with the method of spinor products \cite{13}. As for our
paper,
to analyze the cross-section
of the process $e^+e^- \rightarrow b{\bar b} c{\bar c}$, where the
consideration
for the masses is obligatory,  we apply a more convenient, from our viewpoint,
technique of paper \cite{11}, i.e. the technique of direct numeric calculation
of the amplitude.

Let us introduce the following notations for momenta:
\begin{eqnarray}
q_{12}=q_1+q_2\:, \qquad  p_{12}=&-&p_1-p_2\:,  \qquad  p_{34}=-p_3-p_4\:,
                                                           \nonumber\\
p_{124}=p_{12}-p_4\:, ~~ p_{123}=-p_{12}+p_3\:, \! & & \!
p_{342}=p_{34}-p_2\:, ~~ p_{341}=-p_{34}+p_{1}\:, \nonumber
\end{eqnarray}
for currents
\begin{eqnarray}
J^{\mu}_{\gamma} &=&
Q^e \frac{\bar v (q_2) \gamma^{\mu} u(q_1)}{q^2_{12}}\:,  \\
\nonumber  \\
J^{\mu}_Z &=&
\frac{\bar v (q_2) \gamma^{\mu} (v^e_Z-a^e_Z\gamma^5)
u(q_1)}{q^2_{12}-M^2_Z+iM_Z\Gamma_Z}\:,    \\
\nonumber  \\
J^{\mu}_{b} &=&
\frac{\bar u(p_1) \gamma^{\mu} v(p_2)}{p^2_{12}}\:,  \\
\nonumber  \\
J^{\mu}_{c} &=&
\frac{\bar u(p_3) \gamma^{\mu} v(p_4)}{p^2_{34}}\:,
\end{eqnarray}
and also the following subsidiary spinors
\begin{eqnarray}
v_{124} &=& \frac{({\hat p}_{124}+m_c)}{p^2_{124}-m^2_c} {\hat J}_b v(p_4) \:,
\\    \nonumber \\
v_{342} &=& \frac{({\hat p}_{342}+m_b)}{p^2_{342}-m^2_b} {\hat J}_c v(p_2) \:,
\\    \nonumber \\
{\bar u}_{123} &=& {\bar u} (p_3) {\hat J}_b
\frac{({\hat p}_{123}+m_c)}{p^2_{123}-m^2_c}\:,
\\    \nonumber \\
{\bar u}_{341} &=& {\bar u} (p_1) {\hat J}_c
\frac{({\hat p}_{341}+m_b)}{p^2_{341}-m^2_b}\:,
\end{eqnarray}

In terms of the quantities introduced above the amplitude of the process
$e^+e^- \rightarrow b{\bar b}c{\bar c}$ has the form
$M=\sum\nolimits_{i=1}^{4}M_i$, where

\begin{eqnarray}
M_1 &=& {\bar u} (p_3)
[{\hat J}_{\gamma} Q^c + {\hat J}_Z (v^c_Z-a^c_Z \gamma^5)] v_{124}
\\
M_2 &=& {\bar u}_{123}
[{\hat J}_{\gamma} Q^c + {\hat J}_Z (v^c_Z-a^c_Z \gamma^5)] v (p_4)
\\
M_3 &=& {\bar u} (p_1)
[{\hat J}_{\gamma} Q^b + {\hat J}_Z (v^b_Z-a^b_Z \gamma^5)] v_{342}
 \\
M_4 &=& {\bar u}_{341}
[{\hat J}_{\gamma} Q^b + {\hat J}_Z (v^b_Z-a^b_Z \gamma^5)] v (p_2)
\end{eqnarray}

Here $Q^e$, $Q^b$, $Q^c$ and $v^e_Z$, $a^e_Z$, $v^b_Z$, $a^b_Z$, $v^c_Z$,
$a^c_Z$ are
electromagnetic charges and vector and axial coupling constants of
electron, $b$- and $c$-quark with $Z^0$-boson.

The amplitude squared should be calculated as a sum over
$2^6=64$ independent states of fermions. We choose the eigenstates of the
helicity operator $({\vec \Sigma}{\vec p})$ as two independent states of the
spinor $\psi$:

\begin{equation}
({\vec \Sigma} {\vec p})\psi (p,\lambda)=\lambda \psi (p,\lambda).
\end{equation}

To calculate spinor with the specified value of 4-momentum $p$ and
helicity $\lambda=\pm 1$ it is necessary to choose a specific representation.
In our case we choose a spinor (Weyl) representation:

\begin{equation}
\gamma^0 = \left( \begin{array}{cc}
0 & 1 \\
1 & 0
\end{array} \right)
\qquad {\vec \gamma} = \left( \begin{array}{cc}
0               & -{\vec \sigma} \\
{\vec \sigma}   &        0
\end{array} \right)
\qquad {\vec \Sigma} = \left( \begin{array}{cc}
{\vec \sigma}     &        0   \\
      0           &    {\vec \sigma}
\end{array} \right)
\end{equation}

where
\begin{equation}
{\vec \sigma}=\{ \sigma_x , \sigma_y , \sigma_z \}=
\Biggl\{   \left( \begin{array}{cc}
 0 & 1   \\
 1 & 0 \end{array} \right), \quad
\left( \begin{array}{cc}
 0 & -i   \\
 i & 0 \end{array} \right), \quad
\left( \begin{array}{cc}
 1 & 0   \\
 0 & -1 \end{array} \right)   \Biggr\}
\end{equation}

In this representation the spinors of particles, $u(p,\pm)$,   have the
form
\begin{equation}
\nonumber \\  \nonumber  \\
u(p,+)=\frac{1}{  \sqrt{ 2|{\vec p}|(|{\vec p}| + p_z) }  }
\left( \begin{array}{c}
\sqrt{E+|{\vec p}|} \; (|{\vec p}| + p_z) \\
\sqrt{E+|{\vec p}|} \; ( p_x + i p_y) \\
\sqrt{E-|{\vec p}|} \; (|{\vec p}| + p_z) \\
\sqrt{E-|{\vec p}|} \; ( p_x + i p_y)
\end{array} \right)   \nonumber \\  \nonumber \\ \nonumber
\end{equation}

\begin{equation}
\nonumber\\ \nonumber \\
u(p,-)=\frac{1}{  \sqrt{ 2|{\vec p}|(|{\vec p}| + p_z) }  }
\left( \begin{array}{c}
\sqrt{E-|{\vec p}|} \; (-p_x + i p_y) \\
\sqrt{E-|{\vec p}|} \; (|{\vec p}| + p_z ) \\
\sqrt{E+|{\vec p}|} \; (-p_x + i p_y) \\
\sqrt{E+|{\vec p}|} \; (|{\vec p}| + p_z )
\end{array} \right)  \nonumber  \\ \nonumber \\ \nonumber
\end{equation}

The spinors of antiparticles, $v(p,\pm)$, are defined as

\begin{equation}
v(p,\pm)=C u^*(p,\mp)
\end{equation}
where $C=i \gamma^2$ is a charge conjugation matrix.

The explicit form of spinors $v(p,\pm)$ is as follows:
\begin{equation}
\nonumber\\  \nonumber \\
v(p,+)=\frac{1}{  \sqrt{ 2|{\vec p}|(|{\vec p}| + p_z) }  }
\left( \begin{array}{c}
-\sqrt{E+|{\vec p}|} \; (|{\vec p}| + p_z) \\
-\sqrt{E+|{\vec p}|} \; ( p_x + i p_y) \\
\sqrt{E-|{\vec p}|} \; (|{\vec p}| + p_z) \\
\sqrt{E-|{\vec p}|} \; ( p_x + i p_y)
\end{array} \right)  \nonumber  \\  \nonumber \\ \nonumber
\end{equation}

\begin{equation}
\nonumber \\  \nonumber \\
v(p,-)=\frac{1}{  \sqrt{ 2|{\vec p}|(|{\vec p}| + p_z) }  }
\left( \begin{array}{c}
\sqrt{E-|{\vec p}|} \; (-p_x + i p_y) \\
\sqrt{E-|{\vec p}|} \; (|{\vec p}| + p_z ) \\
-\sqrt{E+|{\vec p}|} \; (-p_x + i p_y) \\
-\sqrt{E+|{\vec p}|} \; (|{\vec p}| + p_z )
\end{array} \right)  \nonumber  \\  \nonumber \\ \nonumber
\end{equation}

But if $p_z=-|{\vec p}|$ we put
\begin{equation}
\nonumber \\  \nonumber \\
u(p,+)=
\left( \begin{array}{c}
0 \\
\sqrt{E+|p|} \\
0 \\
\sqrt{E-|p|}
\end{array} \right)
\qquad  \qquad u(p,-)=
\left( \begin{array}{c}
-\sqrt{E-|p|} \\
0 \\
-\sqrt{E+|p|} \\
0
\end{array} \right)   \nonumber \\ \nonumber  \\ \nonumber
\end{equation}

\begin{equation}
\nonumber \\  \nonumber \\
v(p,+)=
\left( \begin{array}{c}
0 \\
-\sqrt{E+|p|} \\
0 \\
\sqrt{E-|p|}
\end{array} \right)
\qquad  \qquad v(p,-)=
\left( \begin{array}{c}
-\sqrt{E-|p|} \\
0 \\
\sqrt{E+|p|} \\
0
\end{array} \right)   \nonumber \\  \nonumber \\ \nonumber
\end{equation}

The colour structure of the process
$e^+e^- \rightarrow b{\bar b}c{\bar c}$ is very simple.
The colour factor $F$ corresponding to this process is
$F=(N^2-1)/4$, where  N=3 is the number of colours.

The amplitude of the process $e^+e^- \rightarrow b{\bar b}c{\bar c}$
presented by the above formulas can easily be written in terms
of  FORTRAN codes. The resultant program is compact and runs fairly
fast: it takes 0.03s of a VAX/VMS CPU time to compute
the square of the amplitude summed over all  polarization states.

We have verified that our two independent programs written for
calculating the square of the amplitude, summed over all helicity states,
hold the test for Lorentz-invariance (boost along the beam axis) and
test for azimuthal invariance ($p_x \rightarrow p_y$ and
$p_y \rightarrow -p_x$).

The integration over phase space of final particles was carried out
by Monte Carlo methods with the help of programs written for this
process, that minimized the spread in weights. The results of both programs
coincide to an accuracy of the computational errors, caused by
Monte Carlo methods.

\section{The cross-section for the production of
$b{\bar b}c{\bar c}$ at the $Z^0$-peak in QCD perturbation theory}

In our numeric calculations we choose the standard set of the
parameters of electroweak theory:
$M_Z=91.17$~GeV, $\sin^2 {\theta_W}=0.23$, $m_b=4.7$~GeV, $m_c=1.4$~GeV.
The QCD coupling constant $\alpha_s$ is set equal to
\begin{equation}
\alpha_s (Q^2)=\frac{12\pi}{(33-2n_f)ln(Q^2/\Lambda^2)},
\end{equation}
where $\Lambda=100$~MeV and for $Q^2=M^2_Z, n_f=5$ we have
$\alpha_s (M_Z)=0.12$ in accordance with \cite{14}.

It is clear from the form of diagrams on Figs.1a-d  that
the main contribution into the cross section for the process
$e^+e^- \rightarrow b{\bar b}c{\bar c}$ will come from the  diagrams
of Figs.1c,d because for these diagrams the virtual gluon, due to the
small value of the $c$-quark mass, approaches closer the mass shell. A
noticeable peak in the region  $M_{c{\bar c}}=2\div 4 m_c$, related to
the contribution from the diagrams of Figs.1c,d,  is seen on Fig.2a in
the distribution over the invariant mass of  $c{\bar c}$-quarks,
$M_{c{\bar c}}=\sqrt{ {( p_c+p_{\bar c} )}^2 }$. However, the distribution
over the invariant mass of $b{\bar b}$-quarks,
$M_{b{\bar b}}=\sqrt{ {( p_b+p_{\bar b} )}^2 }$, has a completely
different form (see Fig.2b). A small peak in the region of small
masses $M_{b{\bar b}}$ is due to the virtual gluon approaching the mass
shell for those diagrams in which the  $b{\bar b}$-pair is
coupled to the $c({\bar c})$-quark (see diagrams of Fig.1a,b). But if the mass
$M_{b{\bar b}}$ of $b({\bar b})$-quarks is large, then the  mass
$M_{c{\bar c}}$ of $c({\bar c})$-quarks is, as a rule, small and only
diagrams of Figs.1c,d "work", leading to a  pronounced peak
in the region of large masses $M_{b{\bar b}}$.

The process $e^+e^- \rightarrow b{\bar b}c{\bar c}$ takes place
in the second order of the QCD coupling constant  $\alpha_s (Q^2)$.
Therefore it is important to make the right
choice of the specific square of the momentum transferred $Q^2$ in the
argument $\alpha_s (Q^2)$. As it has been mentioned above, the characteristic
scale of the running $\alpha_S(Q^2)$ constant may be defined by the
typical gluon virtuality ($M_{c\bar c}^2 \sim 4m_c^2$), however, the b-quark
virtuality may be more greater and of the order of $s$, so that we present
the results for two different choices of $Q^2 = \{4m_c2,\;s\}$.
Table 1 presents the cross-sections for the process
$e^+e^- \rightarrow b{\bar b}c{\bar c}$ versus the total beam energy
$\sqrt {s}$ for these two cases.

\small
\begin{table}
\caption{The cross-section for the process
$e^+e^- \rightarrow b{\bar b}c{\bar c}$ versus the total beam energy
$\protect\sqrt{ s}$ for the two values of the square of the momentum
transferred: $Q^2=s$ and $Q^2=4m^2_c$.
Bracketed is the error (one standard deviation) for the last digital,
caused by Monte Carlo method.}
\begin{center}
\begin{tabular}{|c|c|c|c|c|c|c|}    \hline
 $\sqrt s$ , GeV & 30 & 50 & 70 & 91.17 & 200 & 500    \\   \hline
$\sigma \bigl({\alpha_s}(s)\bigr)$, pb
& 0.0552(4) & 0.0782(9) & 0.192(3)
& 69.7(2)   & 0.061(2)  & 0.014(1)   \\   \hline
$\sigma \bigl({\alpha_s}(4m^2_c)\bigr)$,pb
& 0.1294(9) & 0.217(3)  & 0.593(9)
& 234.2(7)  & 0.254(8)  & 0.073(5)   \\   \hline
\end{tabular}
\end{center}
\end{table}
\normalsize

As is seen from this table, the difference in the definition of
$\alpha_s$ leads to the difference in the cross-section by some times.
For the most important energy range close to the $Z^0$-boson
peak this difference attains a factor of 3.4. Such a strong
dependence on $\alpha_s$  is related to the fact that the cross-section
for the process $e^+e^- \rightarrow b{\bar b}c{\bar c}$ is
proportional to $\alpha^2_s$. Therefore the accurate  experimental
measurement of the cross-section for the process
$e^+e^- \rightarrow b{\bar b}c{\bar c}$ will allow us
to draw a conclusion on the value of $\alpha^2_s$ in this process.

Let us  define the ratio
\begin{equation}
R_{ c{\bar c} }=\frac{\sigma (e^+e^- \rightarrow b{\bar b}c{\bar c})}
{  \sigma (e^+e^- \rightarrow b{\bar b}) }\:.
\end{equation}

According to our calculations, the value of  $R_{ c{\bar c} }$ is
$R_{ c{\bar c} }=0.8\cdot 10^{-2}$ for $Q^2=s$ and
$R_{ c{\bar c} }=2.6\cdot 10^{-2}$ for $Q^2=4m^2_c$.
These values obtained from the exact formulas of QCD perturbation theory
should be compared with the estimate obtained from the Monte Carlo
program HERWIG (version  5.0),
$R_{ c{\bar c} }=(0.8\div 1.8)\cdot 10^{-2}$ \cite{15}.

Figures 3a,b present the distributions over the  variables
$x=2|{\vec p}|/{\sqrt s}$, where $|{\vec p}|$ is the quark
momentum, for the $b$- and $c$-quark, respectively. One should pay
attention to a completely different nature of the spectra for
the $b$- and  $c$-quarks. The $b$-quarks mainly have a large momentum, whereas
$c$-quarks a small one, pointing once again to the dominating nature of
diagrams with fragmentation of $b$-quarks.

Of course, the free $b$- and $c$-quarks are not
observed experimentally because of confinement. In fact they catch the
light quark and transform into $B$- and $D$-mesons.
To estimate the total and differential inclusive cross-section of
$B_c$-meson production in $p{\bar p}$-collisions the authors of
\cite{2} have used  phenomenological model
of the heavy quark fragmentation into hadrons. This model
describes quite accurately $B$- and $D$-meson production.
In this approach the fragmentation function of the $i$-th quark into
meson, consisting of $i$- and $j$-quarks, has the form \cite{15a}
\begin{equation}
D_{ij}(x)=N~x^{-\alpha_i}~(1-x)^{\gamma -\alpha_j},
\end{equation}
where $\alpha_i$ is the leading Regge trajectory intercept, connected
with the quark of the $i$-type, $\gamma=(1\div 1.5)$ and $N$ is the
normalization coefficient:

\begin{equation}
N=\Gamma (2+ \gamma - \alpha_i -\alpha_j )/\Gamma (1+\gamma -\alpha_j)
\Gamma (1-\alpha_i).
\end{equation}
The values of $\alpha_c$ and $\alpha_b$ lie in the range:
$\alpha_c=-(2\div 2.5)$, $\alpha_b=-(8\div 9)$.

Let us take $\alpha_c$=-2.2, $\alpha_b$=-8.5 and $(\gamma -\alpha_j)$=1.
Then
\begin{eqnarray}
& &D_{b\rightarrow B_c}(x)=99.75~x^{8.5}~(1-x),  \\
& &D_{c\rightarrow B_c}(x)=13.44~x^{2.2}~(1-x).
\end{eqnarray}

Figures 3a,b  present the spectra of $B$- and $D$-mesons, which were
obtained by means of convolution of $b$- and $c$-quarks distributions
with the fragmentation functions of $b$- and $c$-quarks into
$B$- and $D$-mesons,  respectively.

The analysis of angular correlations also discovers a different
behaviour for $b\bar b$- and $c\bar c$-pairs (see Fig.4a,b).
$b$- and $c$- quarks emit mainly in opposite directions whereas
$c$- and $\bar c$-quarks tend to be collinear. This can also
be explained by the aforementioned behaviour of the gluon
propagator for small virtualities.

\section{The $B_c$-meson production cross-section}

In this Section we discuss the production of the $S$-wave
states of $B_c$-mesons. As to the study of the $P$-wave levels,
we plan to make it in our next paper.

The diagrams describing the production of $B_c$-mesons
are obtained from the diagrams of Fig.1a-d by combining
two quark lines into mesonic one (see Fig.5a-d).

The underlying assumption of our calculations is that the binding
energy for two quarks, $b,c$, is much less than their masses and,
hence, heavy quarks in the bound state $B_c$ are actually on
the mass shell. In this case the 4-momenta $p_b$ and $p_c$ of
the quarks-constituents of $B_c$-meson are related with 4-momentum $P$ of
$B_c$-meson as follows:
\begin{equation}
 p_b=\frac{m_b}{M}P\:,  \qquad \qquad   p_c=\frac{m_c}{M}P\:,
\end{equation}
where  $M=m_b+m_c$ is the $B_c$-meson mass.
This assumption corresponds to the leading order in the effective
theory of heavy quarks \cite{16}.

Let us also make use of the fact that the projection operators

\begin{eqnarray}
& & \frac{1}{\sqrt 2} \Bigl\{ v(p,+){\bar u}(p,+)-v(p,-){\bar u}(p,-)  \Bigr\}=
\frac{1}{\sqrt 2}({\hat p}-M)\gamma^5
\\    \nonumber \\
& & \frac{1}{\sqrt 2} \Bigl\{ v(p,+){\bar u}(p,+)+v(p,-){\bar u}(p,-)  \Bigr\}=
\frac{1}{\sqrt 2}({\hat p}-M){\hat \varepsilon}^*(p,0)
\\    \nonumber \\
& & v(p,-){\bar u}(p,+)=\frac{1}{\sqrt 2}({\hat p}-M){\hat \varepsilon}^*(p,+)
\\    \nonumber \\
& & v(p,+){\bar u}(p,-)=\frac{1}{\sqrt 2}({\hat p}-M){\hat \varepsilon}^*(p,-)
\end{eqnarray}

where
\begin{eqnarray}
& &\varepsilon^{\mu}(p,0)=\frac{E}{M|{\vec p}|}
\Bigl\{ \frac{{|{\vec p}|}^2}{E}, p_x, p_y, p_z \Bigr\}
\\    \nonumber \\
& &\varepsilon^{\mu}(p,+)=N
\Bigl\{  0, \varepsilon_x, \varepsilon_y, \varepsilon_z  \Bigr\}
\\    \nonumber \\
& &\varepsilon^{\mu}(p,-)=N
\Bigl\{  0, -\varepsilon^*_x, -\varepsilon^*_y, -\varepsilon^*_z  \Bigr\}
\end{eqnarray}

and
\begin{eqnarray}
& &N=\frac{1}{ 2{\sqrt 2}|{\vec p}|(|{\vec p}|+p_z) }
\\    \nonumber \\
& &\varepsilon_x=-{(|{\vec p}|+p_z)}^2+{(p_x+ip_y)}^2
\\
& &\varepsilon_y=-[{(|{\vec p}|+p_z)}^2+{(p_x+ip_y)}^2]i
\\
& &\varepsilon_z=2(|{\vec p}|+p_z)(p_x+ip_y)
\end{eqnarray}
separate the states with the specified value of the total spin $S$
and its projection $S_z$ on axis $z$ of the system described by the spinors
${\bar u}(p)$ and $v(p)$ ($\varepsilon^{\mu}(p,\lambda)$ is the polarization
vector of the vector state with momentum $p$ and helicity $\lambda$).
Then the amplitude of the process  $e^+e^- \rightarrow b{\bar b}c{\bar c}$
corresponding to the diagrams of Fig.5 can be expressed through these
projection operators and the helicity amplitudes
$M_{h{\bar h}}(\lambda_i)$ found in Section 1 as follows:
\begin{equation}
M(\lambda_i)=\frac{\sqrt {2M}}{\sqrt {2m_b} \sqrt {2m_c}} \Psi (0)
\sum_{h,\bar h}^{}P_{h,\bar h}M_{h,\bar h}(\lambda_i)\:,
\end{equation}
where summation is made over the helicity states  $h,\bar h$ of
the quark and antiquark producing $B_c$-meson. The helicities of
the remaining fermions are symbolically noted through $\lambda_i$.
The projection operators $P_{h,\bar h}$ have the following explicit form
($H=h-{\bar h}$) \cite{17}:

\begin{equation}
P_{h,\bar h}=\frac{1}{\sqrt 2}{(-1)}^{ {\bar h}-1/2 }\delta_{H,0}
\end{equation}
for the  $ ^1 S_0$-state,

\begin{equation}
P_{h,\bar h}=|H|+\frac{1}{\sqrt 2}\delta_{H,0}
\end{equation}
for the $ ^3 S_1$-state.

Taking into account that the colour part of
the $B_c$-meson wave function is
$\delta_{ij}/{\sqrt N}$ the colour factor $F$, corresponding to
the process $e^+e^- \rightarrow B_c{\bar b}c$, is $F=(N^2-1)^2/4N^2$.
The value of the wave function at zero, $\Psi (0)$, is calculated in
nonrelativistic potential model and also in the QCD sum rules  \cite{9} and is
related with the decay constant $f_{B_c}$
of the pseudoscalar  $(0^{-})$ $B_c$-meson and the constant $f_{B^*_c}$ of
the vector  $(1^{-})$ $B^*_c$-meson as follows:

\begin{equation}
\Psi (0)=\sqrt{\frac{M}{12}}f_{B_c},
\end{equation}

where
\begin{equation}
f_{B_c}=f_{B^*_c}=570 \mbox{~Β'}.
\end{equation}

The potentials of various types yield actually the same values of
masses of lower states of $B_c$-mesons. For example, the mass of
pseudoscalar $0^{-}$-meson (1$S$-state) is  $M=6.3$~GeV \cite{6}. In this
connection, when calculating the bound states of $B_c$-mesons of
1$S$ levels we take the values of the masses of $b$- and  $c$-quarks
somewhat larger than those obtained during the production of
free $b{\bar b}c{\bar c}$-quarks and equal $m_b=4.8$~GeV and
$m_c=1.5$~GeV. The mass of 2$S$ levels is predicted to be $M=6.9$~GeV \cite{6}.
In this case the masses of $b$- and  $c$-quarks are taken to be still
larger,  $m_b=5.1$~GeV and  $m_c=1.8$~GeV. We would like to note that
the authors of paper \cite{4} do not take into account the effective increase
of the masses of the $b$- and $c$- quarks, constituents  of $B_c$-mesons,
during the transition to higher excited states of $B_c$-meson. However,
this must be done within the frames of the "on mass shell" formalism
described above. The value of the wave function at zero,
$\Psi (0)$ for $2S$-states is $\Psi (0)=275\mbox{ MeV}^{3/2}$ \cite{6}.

Table 2 presents the calculated cross sections for the production of
$B_c$- and  $B^*_c$-mesons including the first excited levels.
\small
\begin{table}
\caption
{The cross section for the production of $B_c$ ($B^*_c$)-mesons
for ${\protect\sqrt s}=M_Z$ and $\alpha_s (Q^2=M^2_Z)=0.12$. Bracketed is
the error (one standard deviation) in the last digital, caused by
Monte Carlo method}
\begin{center}
\begin{tabular}{|c|c|c|c|c|}    \hline
 $n^{2S+1}L_J$   &  $1{}^1S_0$       &
 $1{}^3S_1$      &  $2{}^1S_0$       &  $2{}^3S_1$           \\   \hline
$\sigma$ , pb    &  0.924(1)         &
 1.285(1)        &  0.2371(3)        &  0.3168(3)            \\   \hline
\end{tabular}
\end{center}
\end{table}
\normalsize

By analogy to what has been said in Section 2 regarding  the dependence of
the cross-section for the process $e^+e^- \rightarrow b{\bar b}c{\bar c}$
on the square of the momentum transferred $Q^2$ in   the argument of
the function $\alpha_s (Q^2)$, similar statement for the process
$e^+e^- \rightarrow B_c{\bar b}c$ also leads to an essential uncertainty
of the predictions regarding the cross-section for the $B_c$-mesons production.
In this connection, the cross-sections for the process
$e^+e^- \rightarrow B_c{\bar b}c$ calculated for $Q^2=s=M^2_Z$
can be interpreted as pessimistic. More optimistic predictions for
the number of $B_c$-mesons to be produced can be  obtained if
as the square of the momentum transferred $Q^2$
we take $Q^2=4m^2_c$.
In this case the cross section for the production of
$B_c$-mesons is ${[\alpha_s (4m^2_c)/\alpha_s (s)]}^2=3.4$ times larger.

The above total cross section of $1S$- and $2S$-states, that is equal to
2.76 pb, is clearly a lower estimate in these calculations. Really,
the total number of the bound states of $B_c$-mesons lying below
the production threshold for $B$- and  $D$-mesons is
about 15. All these excited $B_c$-states with probability equal to 1
transform into $B_c$-meson due to cascade radiative decays. Therefore
higher excited states, not only 1$S$ and 2$S$, can make an essential
contribution at least due to their large number. If we take into
account the production of $\bar B_c$-mesons, the  above
cross-section for the production of $B_c$-mesons should be doubled.

Let us define the ratio of the number of the produced
$B_c({\bar B_c})$-mesons, including also the
$B^*_c({\bar B^*_c})$-states calculated above:
\begin{equation}
R_{B_c}=\frac{ \sigma (e^+e^- \rightarrow B_c{\bar b}c) +
               \sigma (e^+e^- \rightarrow {\bar B_c}b{\bar c})  }
{  \sigma (e^+e^- \rightarrow b{\bar b}) }\;.
\end{equation}

According to our calculations, the value of $R_{B_c}$ is
$R_{B_c}=0.6\cdot 10^{-3}$ for  $Q^2=s$ and
$R_{B_c}=2.0\cdot 10^{-3}$ for  $Q^2=4m^2_c$.
These values obtained from the formulas of QCD perturbation theory
should also be compared with
the estimate derived with the help of the Monte Carlo  program HERWIG,
$R_{B_c}=(0.1\div 1.0)\cdot 10^{-3}$ \cite{15}.

The authors of \cite{4} have recently considered single production of
$B_c$-mesons in  $e^+e^-$-annihilation and obtained the analytical
form for the fragmentation function $D(x)$ of $b$-quark into
$B_c^{(*)}$-meson. Really, in the limit of $s \to \infty$, one can neglect
the terms of the order of $M_2/s$ anf higher. Then the differential
cross section is equal to
\begin{equation}
\frac{d\sigma (B_c^{(*)})}{dz} = N^{(*)}\;\frac{z(1-z)^2}{(1-rz)^6}\;
(f_1^{(P,V)}(z)+f_2^{(P,V)}(z)+f_3^{(P,V)}(z))\;, \label{1a}
\end{equation}
where for the pseudoscalar state one has
\begin{eqnarray}
f_1^{(P)}(z) & = & 2+(2-12r)z+(\frac{2}{3}r- \frac{14}{3}r^2-12r^3)z^3+
(r^2+2r^3+2r^4)z^4\;, \nonumber \\
f_2^{(P)}(z) & = & 2z^2 (1-rz)^2\;, \label{2a} \\
f_3^{(P)}(z) & = & 2z (1-rz)\;(2+(1-6r)z+(r+2r^2)z^2)\;, \nonumber
\end{eqnarray}
and for the vector state one gets
\begin{eqnarray}
f_1^{(V)}(z) & = & 2+(-2-4r)z+(3-6r+12r^2)z^2+ \nonumber \\
{}~ & ~ & (2r-6r^2-4r^3)z^3+(3r^2-2r^3+2r^4)z^4\;, \nonumber \\
f_2^{(V)}(z) & = & 6z^2 (1-rz)^2\;, \label{2c} \\
f_3^{(V)}(z) & = & 6z^2 (1-rz)\;(1-2r+rz)\;, \nonumber
\end{eqnarray}
where $r=m_b/M_{B_C}$\footnote{
In the symmetric $cc \bar c \bar c$ and $bb\bar b\bar b$ case, when the
unflavoured mesons are produced ($\psi$, $\eta_c$, $\Upsilon$, $\eta_b$),
the fragmentation function may be obtained from eqs.(\ref{1a}-\ref{2c})
by the substitution $r=1/2$.},
and the calculations have been performed in the covariant gauge, so that
the $f_{1,2}^{(P,V)}(z)$ functions correspond to the squared diagrams
shown on Figs.1c, 1d, and the $f_{3}^{(P,V)}(z)$ function corresponds
to the interference of those diagrams. In the paper \cite{4} the expression
for the fragmentation of the $b$-quark into the vector $B_c^{*}$ state
coincides with the our one. However, the fragmentation function into the
pseudoscalar meson is not correct in ref.\cite{4}.

Note, the scaling results (\ref{1a}-\ref{2c}) are in a good agreement
with the exact perturbative calculations at $\sqrt{s} = m_Z$ (see Figs.6a, 6b).
However, the nonscaling contributions of the order of $M^2/s$ may be
essential for the both low energies and high mass of the meson like in the
case of the $b$-quark fragmentation into the $\Upsilon$-particle at the
$Z$-pole energy.

Figures 6a,b show our fragmentation functions
$D(x)=\frac{1}{\sigma}\frac{d\sigma}{dx}$ of $b$-quark
into $B_c$-mesons, where
$x=2|{\vec p}|/{\sqrt s}$ and $|{\vec p}|$ is
the  $B_c$-mesons momentum, for the production of pseudoscalar
($0^{-}$) and vector ($1^{-}$)  $B_c$-mesons. The fragmentation function
for vector mesons (see Fig.6b) as compared with
that for pseudoscalar ones (see Fig.6a) has a somewhat sharper
peak in the region $x=0.8$ and a slight dip on the left of the
peak.

Using our computer program it is also easy to obtain some other
distributions. Some of them, the most interesting ones, are
presented below.

Figures 7a,b show the momentum distributions of the associative
$c$-quark for the case of the production of pseudoscalar and
vector $B_c$-mesons.   It is easy to understand that they are
related with the distributions of Fig.6a,b by a simple
relationship:  $D(x_{B_c})\simeq D(1-x_c)$.

Figure 8 presents the distributions in the emission angle between
the pseudoscalar $B_c$-meson and $c$-quark. As is seen from this
figure, the overwhelming number of quarks emit at small angles
$(\leq 30-40^o)$ with respect to the direction of the motion
of $B_c$-mesons, or, as it follows from Fig.9, at small transverse
momenta, $\leq 5-10$~GeV, with respect to the same direction.

The distribution over the emission angle of $B_c(B^*_c)$-mesons with
respect to the beam axis (see Fig.10) can be interesting because of the
effect of the acceptance of real detector on the
number of $B_c(B^*_c)$-mesons to be produced.

Figs.8-10 are shown for the case of the production of
pseudoscalar  $(0^{-})$ $B_c$-mesons. The distributions for
the case of the production of vector $(1^{-})$
$B^*_c$-mesons have the same form, differing from the previous cases
in the normalization only.

\section{Discussions}

So, we have calculated in our paper the cross-section  for the production
of four quarks of different flavours,  $b{\bar b}c{\bar c}$, in
$e^+e^-$-annihilation. At the $Z^0$ peak it is
$70\div 234$~pb and depends upon the choice of the square of the
transferred momentum in the argument $\alpha_s (Q^2)$. When
recalculated for  the number of events, it makes up
$R_{ c{\bar c} }=(0.8\div 2.6)\cdot 10^{-2}$ from the
production of $b{\bar b}$ pairs or $(1.2\div 3.9) \cdot 10^{-3}$ from
the total number of $Z^0$-bosons.

Since the total number of $Z^0$-boson expected from all experiments
by the end of 1994 is about  $2\cdot 10^7$, this corresponds to
$(2.4 \div 7.8)\cdot 10^4$ events with the production of
$b{\bar b}c{\bar c}$-quarks. The experimental study of this process
may be considered as a way to verify the QCD predictions in
higher orders of perturbation theory.

However, the process
$e^+e^- \rightarrow b{\bar b}c{\bar c}$ is also interesting from
another viewpoint. The bound states of $b$- and $c$-quarks, i.e. $B_c$-
mesons, predicted by the theory have not yet been discovered experimentally.
Meanwhile, the discovery and study of $B_c$-mesons may
yield valuable information on the behaviour of the quark
potential in the region intermediate with respect to the families of
$J/\psi$- and $\Upsilon$-mesons. Assuming the existence of quark-hadron
duality, one may relate
the cross-section for the production of the $(b{\bar c})$-system
singlet in colour  (the colour factor is $F=(N^2-1)^2/4N^2$)
in the region of small invariant masses, $M_{b{\bar c}}$,
with the cross-sections for the production of the bound states of
$B_c$-mesons as follows:
\begin{equation}
\int \limits_{m^2_0}^{M^2_{thresh}}
\frac{ {d\sigma ( e^+e^- \rightarrow b{\bar b}c{\bar c} )}_{(b{\bar c})-sing} }
{ dM^2_{b{\bar c}} } dM^2_{b{\bar c}}
=\sum_{}^{} \sigma (e^+e^- \rightarrow B_c{\bar b}c)\:, \label{48}
\end{equation}
where $m_0=m_b+m_c \leq M_{b{\bar c}} \leq M_B + M_D +\Delta M= M_{thresh}
(\Delta M \simeq 0.5 \div 1$~GeV).
Taking $m_0=6.1$~GeV as the boundary values of invariant masses and,
for example, $M_{thresh}=8$~GeV, we obtain the estimate for the cross-section
of the $B_c$-meson production, equal to
$2.07(4) \div 6.9(1)$~pb. This estimate should be compared with
a more accurate calculation of the cross-section for the production of
$B_c$-mesons obtained on the basis of the formalism expounded in
Section 3. The cross-section for the production of $B_c$-meson and
of its  first excitations, 1$S$- and 2$S$-states, without consideration
for other, (about 10), states, whose contribution, according to our estimates,
is suppressed as compared with the contribution from
1$S$- and 2$S$-states, is $2.76 \div 9.3$~pb. This is in a reasonable
agreement with the value following from relationship (48) assuming
quark-hadron duality.

The share of events with the production of $B_c$-mesons
is $R_{B_c}=(0.6\div 2.0)\cdot 10^{-3}$
from the production of $b{\bar b}$ pairs or
$(0.9\div 3.0) \cdot 10^{-4}$ from the total number of events.
As expected,  $2\cdot 10^7$ $Z^0$-bosons will be produced
at LEP. This means that the number of events with the production
of $B_c$-bosons, also including those with the production of
$\bar B_c$-mesons, will be $(1.8\div 6.0) \cdot 10^3 $. Of course,
the real number of experimentally reconstructed events will be less
because one should take into account the branchings of $B_c$-meson
decays into specific modes.  As differed from $J/\psi$- and
$\Upsilon$-mesons, $B_c$-mesons do not have decays of annihilation type
into 2 or 3 gluons and decay into a weak channel. The decay
$B_c\rightarrow J/\psi +X$ with a  branching of $\sim$~30\%
and further decay of $J/\psi$ into two muons with a branching of 6\%
is the most
promising in terms of experimental identification of the final
$30\div 110$ reconstructed events with the production of
$B_c({\bar B_c})$-mesons. $B_c$-mesons have quite a long
lifetime and, hence, a long decay length, that may be
helpful for identifying the vertex of $B_c$-meson decay.
The availability of vertex detectors and separation of hard leptons
in the jets from  the decays of    $b$- and  $c$-quarks in the
process $e^+e^- \rightarrow B_c{\bar b}c$ might also be helpful
for identifying the final state, specific for the process with
the production of $B_c$-mesons. However, one should take into
account that this channel of $B_c$-meson decay has a noticeable
background coming from the decays of ordinary (noncharmed) rather than charmed
$B$-mesons into $J/\psi$-mesons. Therefore a thorough analysis of
this background is required for separation of the events with the
production of $B_c$-mesons.

It is interesting to  note that the cross section of the production
of a  (bc)-pair with an invariant mass of a
$M_{bc}$  pair in the region  $m_0 \leq M_{bc} \leq M_{thresh}$
(see formula (\ref{48}))
is also a few $pb$. It is natural to assume that such a pair of $bc$-quarks,
catching a light quark can fragment with a high probability
into a colourless object, $\Lambda_{bc}$-hyperon,
decaying further in a cascade way. According to our estimates,
there should be some thousands of such events per $2\cdot 10^7$ $Z^0$-
bosons. Observation of new hyperon $\Lambda_{bc}$ and of $B_c$-meson
is an interesting and quite realistic problem of the nearer future.

\vspace*{0.cm}
\section*{Acknowledgments}
In conclusion, the authors  would like to express their sincere
gratitude to E.R.~Budinov, A.A.~Likhoded, A.V.~Tkabladze, S.A.~Shichanin,
and O.P.~Yu\-sh\-chen\-ko for helpful discussions and valuable remarks.

{\hfill \small  Received April 22, 1993}

\begin{figure}[p]
\caption{Feynmann diagrams for the process
$e^+e^- \rightarrow b{\bar b}c{\bar c}$.}
\end{figure}
\begin{figure}[p]
\caption{a) Distribution over invariant mass
$M_{c{\bar c}}=\protect\sqrt{ {( p_c+p_{\bar c} )}^2 }$ of
$c{\bar c}$-pair.
b) the same as in a) but for
$M_{b{\bar b}}=\protect\sqrt{ {( p_b+p_{\bar b} )}^2 }$ of  $b{\bar b}$-pair.}
\end{figure}
\begin{figure}[p]
\caption{a) Distributions over
$x=2|{\vec p}|/{\protect\sqrt s}$, where $|{\vec p}|$ is the $b$-quark momentum
(solid line) or $B$-meson momentum (dotted line).
b) The same as in a),  where $|{\vec p}|$ is the $c$-quark momentum
(solid line) or $D$-meson momentum (dotted line).}
\end{figure}
\begin{figure}[p]
\caption{a) Distribution over the angle between the directions of motions of
$b$- and $\bar b$-quarks.
b) The same as in a) but for $c$- and $\bar c$-quarks.}
\end{figure}
\begin{figure}[p]
\caption{Feynmann diagrams for the process
$e^+e^- \rightarrow B_c{\bar b}c$.}
\end{figure}
\begin{figure}[p]
\caption{a) Fragmentation function of pseudoscalar $B_c$-meson
vs. the variable $x=2|{\vec p}|/{\protect\sqrt s}$, where $|{\vec p}|$ is
$B_c$-meson momentum.
b) The same as in a) but for vector $B^*_c$-meson.}.
The curves are the scaling expressions (\ref{1a}-\ref{2c})
\end{figure}
\begin{figure}[p]
\caption{a) Distribution over variable
$x=2|{\vec p}|/{\protect\sqrt s}$, where $|{\vec p}|$ is
$c$-quark momentum for the case of the pseudoscalar $B_c$-meson production.
b) The same as in a) but for the vector $B^*_c$-meson production. }
\end{figure}
\begin{figure}[p]
\caption{Distribution over the angle between the directions of motions of
$B_c$-meson and $c$-quark.}
\caption{Distribution over the transverse momentum of $c$-quark relatively
to the direction of motion of $B_c$-meson.}
\end{figure}
\begin{figure}[p]
{\caption{Distribution over the angle of $B_c$-meson with respect to
the $e^-$-beam direction.}}
\end{figure}
\end{document}